\documentclass{article}
\usepackage{spconf,amsmath,graphicx}
\usepackage{multirow,makecell,hhline}
\usepackage[normalem]{ulem}
\usepackage{booktabs}
\usepackage{tikz}
\usepackage{cite}
\usepackage{url}
\usepackage{hyperref}
\hypersetup{
    colorlinks=true,
    linkcolor=red,
    urlcolor=magenta,
    }

\newcommand{\tbh}[1]{\textbf{#1}}

\newcommand{\tabincell}[2]{\begin{tabular}{@{}#1@{}}#2\end{tabular}}

\title{UML: A Universal Monolingual Output Layer for Multilingual ASR}
\name{Chao Zhang$^{\dagger\ddagger}$, Bo Li$^{\dagger}$, Tara N. Sainath, Trevor Strohman, Shuo-yiin Chang\thanks{$\dagger$Equal contributions. $\ddagger$Work performed while at Google.}}
\address{Google LLC, USA\\
\small{\texttt{\{boboli,tsainath,strohman,shuoyiin\}@google.com}}}
\begin{document}
\ninept
\maketitle
\begin{abstract}
Word-piece models (WPMs) are commonly used subword units in state-of-the-art end-to-end automatic speech recognition (ASR) systems. For multilingual ASR, due to the differences in written scripts across languages, multilingual WPMs bring the challenges of having overly large output layers and scaling to more languages. In this work, we propose a universal monolingual output layer (UML) to address such problems. Instead of one output node for only one WPM, UML re-associates each output node with multiple WPMs, one for each language, and results in a smaller monolingual output layer shared across languages. Consequently, the UML enables to switch in the interpretation of each output node depending on the language of the input speech. Experimental results on an 11-language voice search task demonstrated the feasibility of using UML for high-quality and high-efficiency multilingual streaming ASR. 
\end{abstract}

\section{Introduction}
\label{sec:intro}
Automatic speech recognition (ASR) is used by a massive amount of users. Although more than 7,100 languages are actively spoken in the world \cite{Ethnologue}, only about a hundred most common ones have commercial ASR products, restricting the benefits of such useful artificial intelligence technology for people. To extend ASR to cover more languages and users \cite{Slaney2020}, it is better to serve many languages with a single multilingual ASR system instead of with many monolingual ASR systems, which not only enables code-switch and resource sharing across languages but also reduces maintenance cost.

The choice of subword units is critical for multilingual ASR. 
Phonemes are often used in the modularised ASR framework \cite{Knill2013,ZhangC2014,Vu2014,LiX2020}, which requires complex text processing to convert pronunciations to words. Text units are often used to resolve the issue. 
Graphemes, as a collection of characters, can result in an uneven distribution of subword units across languages \cite{Watanabe2017,Kim2018,Hou2020,Toshniwal2018}. 
As an alternative, bytes can be used as the common subword units shared across languages by decomposing each character into multiple bytes \cite{LiB2019}.
Word-piece models (WPMs) \cite{Schuster2012,Sennrich2016} and similarly sentence-piece models \cite{Kudo2018}, obtained by segmenting words or sentences into pieces, are superior in performance to graphemes and bytes \cite{Rao2017,Zeyer2018} and are therefore the \textit{de facto} choices in monolingual end-to-end ASR and natural language processing.
However, it is inevitable to have a large number of multilingual WPMs when multiple writing systems are involved \cite{LiB2021,Zhou2022,Pratap2020,Pham2022}.  
A solution is to use separate monolingual output layers and decoders \cite{Joshi2021,Mavandadi22}, which considerably increases the storage space and requires the management of more concurrent beam search procedures when scaling up to more languages.

In this paper,  we propose a universal monolingual output layer (UML) to resolve the aforementioned issues. Compared to using separate monolingual output layers or decoders, the UML reuses the same output layer for different languages, simply by re-associating each output node with different subword units for different languages. This can be achieved with only a  tokeniser-level change, which allows a multilingual decoder to keep almost the same structure and size as the monolingual decoder.
Compared to graphemes and bytes, better performance is expected as monolingual WPMs with controllable sizes can be used in UML.  
Streaming ASR experiments on a massive multilingual voice search setup \cite{LiB2021} with 11 languages and known language identifiers (LIDs) showed that compared with the baseline with 8K output nodes, UML with four thousand (4K) output nodes and a rule-based language grouping achieved the same word error rate (WER) by reducing about 40\% of parameters in decoders. 
A further reduction of the decoder size was achieved by replacing WPMs with bytes for logogram-coded languages (\textit{e.g.} Chinese and Japanese), which resulted in better WERs than a byte-only baseline with an output layer size of 512. 

In the rest of the paper: Sec.~\ref{sec:relatedwork} reviews related work. Sec.~\ref{sec:methods} presents the proposed UML method. Sec.~\ref{sec:exp_setup} and~\ref{sec:results} are the experimental setup and results. We conclude in Sec.~\ref{sec:conclusions}.

\section{Related Work}
\label{sec:relatedwork}
Phonetic units, including shared and language-dependent phonemes, decision-tree clustered context-dependent phonemes \cite{Knill2013,Vu2014,ZhangC2014,LiX2020}, and articulatory features \cite{ZhangC2011}, are commonly used in modularised ASR. Although it is natural to model the pronunciation variations across languages with phonetic units, the need for lexicons and language models made it less suitable for on-device applications.



Context-dependent graphemes were first developed for modularised ASR. 
Decision tree clustering is often used to cover unseen graphemes and complex segmental writing scripts, whose question sets can be derived based on the Unicode character descriptions \cite{Gales2015}.
Context-independent graphemes are prevalent in end-to-end ASR, such as recurrent neural network transducer (RNN-T). 
For multilingual ASR, a union of monolingual graphemes is used \cite{Watanabe2017,Kim2018,Hou2020,Toshniwal2018}.
By segmenting each grapheme into one to four Unicode bytes based on its UTF-8 encoding, the same bytes can be used to build ASR to transcribe any writing scripts \cite{LiB2019}. Having a small number of common ``sub-character'' units, byte ASR models require more steps to decode than graphemes and the partial outputs are not human-readable. 

By incrementally combining graphemes into frequent pieces of words, the vocabulary size can be controlled using WPMs \cite{Schuster2012}. Since fewer decoding steps can be achieved with more WPMs, with a sufficient number of WPMs, RNN-T models were found not only to produce better WERs than with graphemes \cite{Rao2017}, but also to produce similar WERs using a full history decoder and a limited history decoder \cite{prabhavalkar2021less}. As graphemes can be considered as the minimum WPMs, in practice, WPMs and graphemes are often mixed to use in multilingual ASR. For instance, similar numbers of English WPMs and Chinese characters are often used together in  English-Chinese code-switch studies \cite{Zeng2019,Dalmia2021}, to avoid the 26 English characters being overwhelmed by thousands of Chinese characters. 
As an extension of WPMs, the sentence-piece model allows cross-word subword segmentation based on different algorithms \cite{Kudo2018}. 


Although WPMs are more commonly adopted over graphemes for monolingual ASR, an output layer with multilingual WPMs generated by pooling all monolingual data together can often be overly large when many languages and writing scripts are integrated \cite{adams2019massively,li2022massively}. Separate monolingual output layers can be used as a solution, which can be dated back to the previous works with phonemes and graphemes \cite{Heigold2013,Huang2013}. In RNN-T, other parts of the decoder and even a part of the encoder can be monolingual as well \cite{Joshi2021,Mavandadi22}. The UML method proposed in this paper differs from all these works by tying all monolingual output layers together, which enables the monolingual ASR decoder structure to be used for multilingual ASR. 
It is worth noting although UML is introduced for WPMs, it is a generic method applicable to other kinds of subword units as well.  

\section{Proposed UML Method}
\label{sec:methods}

\subsection{A Universal Monolingual Output Layer}
\label{ssec:uml}
UML is a monolingual output layer shared by all languages. 
Specifically, let $L$ be the number of languages, $V_l$ be the number of WPMs for the $l$-th language, in UML, each output node $o$ is mapped to $L$ different monolingual WPMs $W_{1,o},
\ldots,W_{L,o}$ for $L$ different languages, whereas $o$ is mapped to only one WPM in a conventional output layer. 
Let $H$ be the input dimension of the output layer, compared to alternative methods, UML enables the use of more WPMs with fewer parameters:



\begin{itemize}
\item UML uses only one $H\times\max(V_1,\ldots,V_L)$-dimensional (-dim) output layer to model the $\sum\nolimits_{l=1}^LV_l$ WPMs.

\item The method using a conventional output layer for all multilingual WPMs \cite{Punjabi2021,LiB2021,Zhou2022,Pratap2020,Pham2022} requires to use a $H\times(\sum\nolimits_{l=1}^LV_l)$-dim layer for the $\sum\nolimits_{l=1}^LV_l$ WPMs. 

\item The methods in \cite{Joshi2021,Mavandadi22} use $L$ separate monolingual output layers whose dimensions are $H\times V_1,\ldots,H\times V_L$. It requires $H\times(\sum\nolimits_{l=1}^LV_l)$ parameters to model the $\sum\nolimits_{l=1}^LV_l$ WPMs. 

\end{itemize}


In UML, since each WPM is determined jointly by the LID and the output node index, LIDs need to be taken into account in inference.  
At test-time, let $\mathbf{x}$, $\mathbf{y}$, and $\mathbf{z}$ be the input, output, and LID prediction sequences of an utterance, $\mathbf{y}^{*}$ is the decoding result, the \textit{maximum a posteriori} decoding rule of ASR can be modified to marginalise the LID predictions as
\begin{align}
    \label{eq:uml}
    \mathbf{y}^{*}=\arg\max P(\mathbf{y}|\mathbf{x})=\sum\nolimits_{\mathbf{z}} P(\mathbf{z}|\mathbf{x})P(\mathbf{y}|\mathbf{x},\mathbf{z}),
\end{align} 
where $P(\mathbf{y}|\mathbf{x},\mathbf{z})$ are the output distributions of a UML-based LID-aware multilingual decoder and a LID predictor, $P(\mathbf{z}|\mathbf{x})$ are the LID predictions.
Eqn.~\eqref{eq:uml} can also be applied to the training, which has $P(\mathbf{y}|\mathbf{x})$ in the loss function. 
This allows LID prediction and code-switch to be handled jointly and explicitly. 

There are a few key advantages to using the UML: \begin{itemize}
 \item First, the UML allows multilingual ASR to scale gracefully to any number of languages without increasing the output layer size \cite{adams2019massively,li2022massively}. This is smaller in size than the conventional multilingual output layer and improves the computation efficiency in both RNN-T training and decoding. It also reduces the difficulty to adapt an existing system to new languages by reusing the same output layer. 

\item Second, the UML provides us with fine-grained control of WPM tying across languages.
Languages with similar writing systems (\textit{e.g.} English and French) often have duplicated WPMs. 
To avoid this, such languages can be grouped to derive a combined set of WPMs. In this setting, the UML is no longer fully ``monolingual'' by tying the same WPMs within each group. Meanwhile, the flexibility is reserved to untie some WPMs with the same written form but dissimilar pronunciations and contexts, such as the same characters in Chinese and Japanese, by setting them into different groups.

\item Third, the UML enables us to handle contextual biasing in a per-language way. Due to a page limit, contextual biasing is not included in this paper, but it is necessary for real-world ASR applications that can require us to incorporate thousands of biasing words (such as contact names) into ASR. In multilingual ASR, each UML output distribution is ``monolingual'' enough to keep the relevant biasing word list monolingual, without the need to combine them into an overly large list.

\end{itemize}


\subsection{Applying UML to Multilingual RNN-T}
\label{ssec:uml}

\begin{figure}[t]
    \centering
    \includegraphics[scale=0.45]{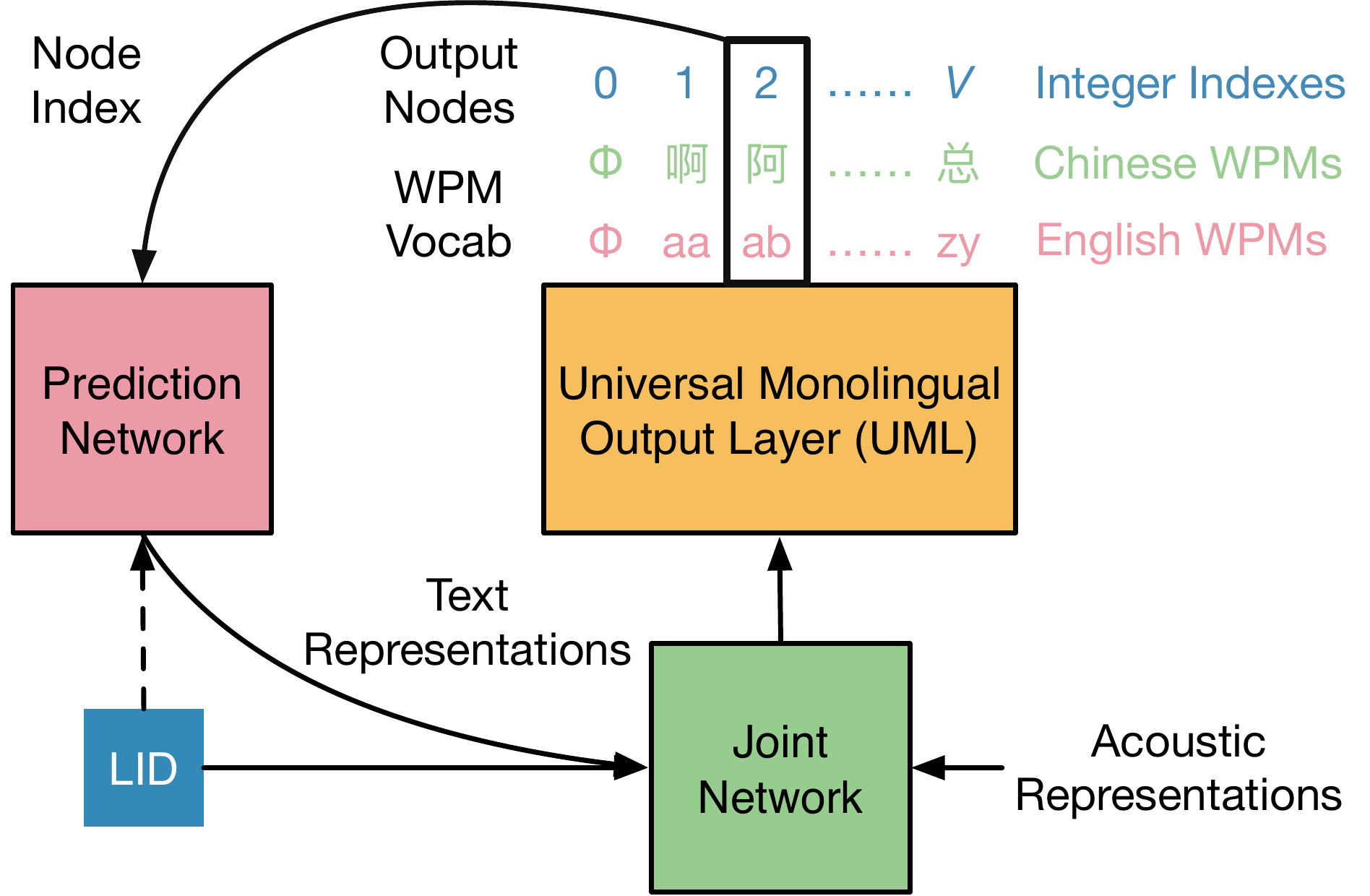}
    \caption{A sketch map of an RNN-T decoder with WPM-based UML for English and Chinese. LID to the prediction network is optional.}
    \label{fig:uml}
\end{figure}

In this section, the application of UML in an RNN-T is considered. As shown in Fig.~\ref{fig:uml}, the output layer of an RNN-T decoder produces its current output distribution based on the joint representation derived from the joint network. The standard joint network is a single hidden layer in RNN-T, which fuses the acoustic representation derived from the encoder based on the input speech, with the text representation derived from the prediction network based on the outputs from the previous time steps. Not only the output layer but also the prediction network needs to be considered when using UML. 

Since LID is missing from the prediction network, the input to the prediction network, which is the output node index of the previous step, is ambiguous as it can not determine a WPM alone in UML. 
To disambiguate, the LID information can be leveraged in the prediction network, by either augmenting each output node index with a LID or using language-specific parameters. 
As shown in Eqn.~\eqref{eq:uml}, this requires expanding the search space in decoding, by re-weighting each path in the beam with the relevant LID probability. 
Alternatively, if LID is ignored in the prediction network, the LID predictor, such as the one proposed in \cite{ZhangC22b}, can be decoupled from the beam search that simplifies UML implementation. 

In the applications where per-utterance oracle LIDs are available, $P(\mathbf{z}|\mathbf{x})$ are 0-1 hard probabilities and Eqn.~\eqref{eq:uml} becomes $\mathbf{y}^{*}=P(\mathbf{y}|\mathbf{x},\mathbf{z})$. Using UML in this case does not require any change to model training and testing. Only the output node indexes are interpreted according to the language, which can be achieved simply by switching among a set of monolingual tokenisers controlled by the LIDs. Here a tokeniser converts a sequence of output node indexes into a sequence of WPMs and joins them into word-level transcriptions, or \textit{vice versa}. As a result, a UML-based multilingual ASR system can have the same decoder structure and size as a monolingual ASR system, which results in lower storage, computation, and energy costs. In this paper, we focus on UML with oracle LIDs.

In UML, having a different number of WPMs for different languages requires us to apply partial matrix multiplications and softmax functions with different sizes. For simplicity, we unify all sets of WPMs to have the same size $V$. The UML in this case can be viewed as folding the multilingual decoder output layer $L$ times. Furthermore, to reduce the output layer size to be smaller than the number of Chinese characters, bytes can be used as the alternative subword units for Chinese and Japanese, which have a different vocabulary size than those of the WPMs used for other languages.

\section{Experimental Setup}
\label{sec:exp_setup}

\subsection{Data}

Our dataset consists of 11 languages: Arabic, Chinese, German, English, Spanish, French, Hindi, Italian, Japanese, Portuguese and Russian. For each language, the training data is collected from two application domains: Voice Search and YouTube, with complex noise and acoustic conditions. All data are de-identified and human transcribed. All data collection and handling abide by Google AI Principles \cite{googleaiprinciples}. Table \ref{table:data} shows the distribution in more detail.
Each test set utterance is less than 5.5 seconds in length and is sampled from the traffic of Google's Voice Search product. All test sets are similarly de-identified and human transcribed for evaluation. All data are normalised to have zero mean and unit variance based on the statistics collected from the whole training set.

\begin{table}[t]
\centering
\vspace{-0.1in}
\begin{tabular}{llrrr}
\toprule
\multirow{2}{*}{\tbh{LID}} & \multirow{2}{*}{\tbh{Language}} & \tbh{\#Utterance} & \tbh{Duration} & \tbh{\#Utterance} \\
 &  & \tbh{Train (M)} & \tbh{Train (K)} & \tbh{Test (K)} \\
\midrule
\midrule
ar & Arabic & 3.8 & 6.4 & 4.1\\
de & German &  3.8 & 3.8 & 7.6\\
en & English & 18.1 & 15.7 & 9.8\\
es & Spanish & 45.5 & 52.8 & 17.5\\
fr & French & 10.8 & 14.7 & 7.1\\
hi & Hindi & 14.2 & 29.8 & 6.5\\
it & Italian & 13.0 & 21.3 & 14.8\\
ja & Japanese & 10.9 & 11.5 & 11.2\\
pt & Portuguese & 13.4 & 20.7 & 12.3\\
ru & Russian & 5.3 & 12.4 & 11.1\\
zh & Chinese & 0.9 & 5.1 & 6.1\\
\midrule 
\midrule
Total & & 135.0 & 182.6 & 108.1 \\
\bottomrule
\end{tabular}
\caption{Data statistics. The number of utterances (\#Utterance) is in millions (M) and the duration is in thousands (K) of hours.}
\label{table:data}
\end{table}

\vspace{-3pt}
\subsection{Model}
\vspace{-1pt}

The Conformer-based cascaded encoder RNN-T model is used in this study \cite{Narayanan2021}. 
The acoustic features are 128-dim log Mel filter banks computed on a 32ms window with a 10ms shift. SpecAugment is used to improve model robustness. 
Every 4 contiguous frames with 3 frames on the left are stacked to form a 512-dim input representation, which is further sub-sampled with one frame overlap to achieve a 30ms frame rate. Oracle LID vectors are one-hot-coded and appended to the acoustic features. 
The causal encoder has 46 million (M) parameters and consists of 3 convolution layers followed by seven Conformer blocks. Causal convolution and left-context attention layers are used in the Conformer block to strictly exclude any right context frame for streaming purposes. 
The right-context encoder with 100M parameters uses a 640-dim linear projection to transform the input features, followed by ten 640-dim Conformer blocks and a final linear normalisation layer. 
More details about the encoder structures and training configurations are in \cite{ZhangC22a}.

Though separate decoders with identical structures are used in all experiments, only WERs from the 2nd-pass decoder are reported. Each decoder has an embedding prediction network \cite{prabhavalkar2021less} operating on 2 non-blank model predictions with 640-dim embeddings, which is termed as an \textit{embedding decoder}. For the output layer size $O$, there are two 640$\times O$-dim embedding layers. The two 640-dim embeddings for the previous two non-blank predictions are concatenated and projected down to a 640-dim text representation, resulting in a 1280$\times$640-dim linear projection. Two projections are used to further transform the encoder and prediction network outputs separately.   
The joint network fuses these transformed representations into a 640-dim vector and then passes it to the final 640$\times O$-dim softmax output layer. 
The total number of parameters for each decoder is 18.1M when $O$ is 8K. A decoder structure with an LSTM prediction network is evaluated for comparison, which has 26.5M parameters. The LSTM prediction network consists of two projected LSTM layers, whose memory cell and hidden representations are 1280-dim and 640-dim.  A bilinear pooling (BP) joint network, referring to the Eqns. (\textcolor{red}{11}) and (\textcolor{red}{12}) in \cite{ZhangC22a}, can be used to improve the decoder performance, which increases the joint network size by 1.1M parameters. 
The LID information is not used in the prediction network of the UML systems unless claimed explicitly. 


\section{Results and Discussions}
\label{sec:results}
\begin{table*}[!ht]
\caption{Comparisons on test set \%WERs and numbers of decoder parameters (Decoder \#Params) in million (M). The ``B'' and ``U'' systems are the baseline and UML systems. ``512'', ``1k'', ``2k'', ``4k'', ``6k'' and ``8k'' are the output layer sizes. The ``G''s are followed by a number of distinct languages or groups used in UML, and ``G5Mix'' replaces the WPMs of Chinese and Japanese with bytes in G5 in UML.  ``LSTM'', ``BP'', and ``LID'' refer to the use of the LSTM prediction network, the BP joint network, and the LIDs in the prediction network respectively. }
\centering
\resizebox{0.95\textwidth}{!}{
\begin{tabular}{lrrrrrrrrrrrrr}
\toprule 
\multirow{2}{*}{\tbh{System}} & \multirow{2}{*}{\tbh{\tabincell{c}{Decoder \\\#Params}}} & \multirow{2}{*}{\tbh{\tabincell{c}{Mean \\\%WER}}} & \multicolumn{11}{c}{\tbh{Per-language \%WERs}}\\
\cmidrule(lr){4-14}
~ & ~ & ~ & \tbh{ar} & \tbh{de} & \tbh{en} & \tbh{es} & \tbh{fr} & \tbh{hi} & \tbh{it} & \tbh{ja} & \tbh{pt} & \tbh{ru} & \tbh{zh}  \\
\midrule
\midrule
B0$^\text{8kG1}$ & 18.1M & \textbf{11.3} & 11.7  & 13.2 & 7.8 & 6.5 & 9.8 & 19.9 & 9.3 & 14.2 & 7.7 & 12.3 & 11.4 \\
B1$^\text{8kG1}_\text{LSTM}$ & 26.5M & \textbf{10.9} & 10.8  & 12.9 & 7.6 & 6.1 & 9.6 & 19.8 & 9.0 & 13.3 & 7.5 & 11.7 & 11.9 \\
B2$^\text{Bytes}$ & 3.1M & 12.6 & 12.3 & 14.1 & 8.8 & 7.2 & 11.1 & 22.4 & 10.5 & 17.1 & 8.9 & 14.3 & 11.4 \\
\midrule
U0$^\text{6kG11}$ & 14.2M & 12.0 & 12.4 & 14.0 & 8.1 & 7.2 & 10.6 & 20.2 & 10.5 & 16.0 & 8.4 & 13.1 & 11.9\\
U1$^\text{6kG7}$ & 14.2M & 11.7  & 11.7 & 13.6 & 8.0 & 6.9 & 10.1 & 20.0 & 9.8 & 15.8 & 8.0 & 12.7 & 11.7 \\
U2$^\text{6kG5}$ & 14.2M & 11.7 & 11.5  & 13.2 & 7.9 & 6.5 & 10.0 & 20.0 & 9.7 & 16.6 & 8.1 & 12.8 & 12.0\\
U3$^\text{6kG6}$ & 14.2M & 11.5 & 11.6  & 13.2 & 7.9 & 6.5 & 9.9 & 19.7 & 9.6 & 15.3 & 8.0 & 12.8 & 11.8 \\
U4$^\text{6kG6}_\text{BP}$ & 15.3M & \textbf{10.9} & 11.1 & 12.3 & 7.1 & 6.2 & 9.4 & 19.5 & 9.0 & 13.9 & 7.5 & 12.0 & 11.9 \\
\midrule
U5$^\text{4kG6}$ & 10.2M & 11.7 & 11.7  & 13.5 & 7.8 & 6.6 & 10.0 & 19.8 & 9.8 & 15.6 & 8.0 & 12.9 & 13.1 \\
U6$^\text{4kG6}_\text{BP}$ & 11.3M & \textbf{11.3} & 10.9 & 12.1 & 7.3 & 6.1 & 9.4 & 19.4 & 9.0 & 17.2 & 7.4 & 11.8 & 13.5 \\
U7$^\text{4kG6}_\text{LID}$ & 10.2M & 11.7 & 11.5  & 13.3 & 7.8 & 6.6 & 10.1 & 19.8 & 9.7 & 15.6 & 7.9 & 12.9 & 13.3 \\
U8$^\text{2kG6}$ & 6.3M & 12.5 & 11.3  & 13.5 & 7.9 & 6.5 & 10.0 & 19.9 & 9.9 & 16.8 & 7.9 & 12.9 & 21.2 \\
\midrule
U9$^\text{1kG5Mix}_{\text{BP}}$ & 4.7M & 11.9 & 11.9  & 13.8 & 8.4 & 6.9 & 10.4 & 20.2 & 10.2 & 16.0 & 8.4 & 13.7 & 11.3 \\
U10$^\text{512G5Mix}_{\text{BP}}$ & 3.7M & 12.1 & 11.8  & 14.3 & 8.3 & 6.9 & 10.6 & 20.4 & 10.3 & 16.4 & 8.7 & 14.0 & 11.5 \\
\bottomrule
\end{tabular}}
\label{tbl:results_all}
\end{table*}

Full ASR results are presented in Table~\ref{tbl:results_all} and discussed as follows.

\textbf{Baseline models (B0--B2).} B0$^\text{8kG1}$ is a baseline system using a single conventional output layer with 8K multilingual WPMs. The multilingual WPMs were generated by pooling the training data of all languages into one group denoted as G1. 
We use the natural distribution of each language in WPM generation as well as the mini-batch sampling in ASR training. 
For on-device applications, the smaller decoder is always preferred for better efficiency and latency. In our 8K vocabulary embedding decoder, most parameters are in the output layer and embedding look-up tables. The 8K WPM set is the smallest we can build without bringing many out-of-vocabulary (OOV) tokens due to the reduction of Chinese characters. B1$^\text{8kG1}_\text{LSTM}$ is another baseline with an LSTM prediction network. 
Our third baseline B2$^\text{Bytes}$ has 384 output nodes including 256 for bytes and 128 for special tokens. Due to the small decoder size, B2$^\text{Bytes}$ has much worse WER than B0$^\text{8kG1}$ and B1$^\text{8kG1}_\text{LSTM}$, apart from Chinese.

\textbf{UML feasibility and language grouping (U0-U4).} To test the feasibility of the UML approach, we start with systems U0--U4, whose structures are similar to B1 but have a 6K UML instead of an 8K conventional output layer. 6K is the least WPMs for Chinese and Japanese on our data without causing OOVs. U$0^\text{6kG11}$ has 11 distinct sets of monolingual WPMs without any grouping. G7 groups the Germanic (de and en) and the Romance (es, fr, it, and pt) languages separately, and hence U$1^\text{6kG7}$ has 7 distinct languages or groups. To combine all languages with similar writing scripts, G5 further merges the Germanic and Romance languages into one group, and Chinese and Japanese into another group, leading to U$2^\text{6kG5}$ with 5 distinct language groups. From the results, both U$1^\text{6kG7}$ and U$2^\text{6kG5}$ have lower WERs than U$0^\text{6kG11}$, which are still worse than B0$^\text{8kG1}$. This proves that building a UML-based system without any grouping is difficult. 
U$2^\text{6kG5}$ were worse than U$1^\text{6kG7}$ in Chinese and Japanese, but outperformed in other languages, indicating that even the same characters in Chinese and Japanese can have very different pronunciations and contexts that are not suitable to share the same nodes. Therefore, we built another group G6, by splitting Chinese and Japanese from G5, and the resulting system U$3^\text{6kG6}$ indeed had improved WERs in these two languages. Such fine-grained control is a benefit of using UML. Further replacing the joint network in U$3^\text{6kG6}$ with BP with a small parameter quantity increase, resulted in our best-performing UML system U$4^\text{6kG6}_\text{BP}$, which has the same WER as our best LSTM-based baseline B1$^\text{8kG1}_\text{LSTM}$ with much fewer parameters. U$4^\text{6kG6}_\text{BP}$ outperformed B0$^\text{8kG1}$ by relatively 3.5\% lower WER and  reduced 15\% of the parameters. 

\textbf{Reducing WPM-based UML size to the limit (U5-U8).}
Next, we further reduced the size of all G6 WPM sets to 4K and 2K. U5$^\text{4kG6}$ have a 0.2\% increase in the mean WER compared with U3$^\text{6kG6}$, while most of the increase was caused by the OOVs in Chinese. By further using the BP joint network, U6$^\text{4kG6}_\text{BP}$ has another 3.5\% relative WER reduction, and the result is the same mean WER as B0$^\text{8kG1}$ while saving about 40\% of the decoder parameters. Meanwhile, we tested the importance of including the LID information in the prediction network. Two additional $11\times 640$-dim projections are used to transform the LIDs and their outputs are added to the 640-dim embeddings of the previous two steps accordingly. The resulting system U7$^\text{4kG6}_\text{LID}$ had the same mean WER as U5$^\text{4kG6}$, meaning it is not necessary to disambiguate the output node indexes with the LID information in this case. 
Furthermore, a 2K WPM-based UML system, U8$^\text{2kG6}$ is built, with even more OOVs in Chinese and Japanese than U5$^\text{4kG6}$. The mean WER is much worse than U5$^\text{4kG6}$ as expected, but the WER increases are mostly in Chinese and Japanese. This inspires us to further improve the method by using bytes for these two languages. 

\textbf{Using UML to mix WPMs and Bytes (U9, U10).} To avoid the degradation of WERs due to OOV issues, we propose to use bytes for the logogram-coded languages, WPMs for other alphabet-coded languages and mix such different types of subword units using UML. The resulting 1K output-node and 512 output-node systems with BP joint network, U9$^\text{1kG5Mix}_{\text{BP}}$ and U10$^\text{512kG5Mix}_{\text{BP}}$, both achieved better WERs with fewer parameters than U8$^\text{2kG6}$. In addition, U10$^\text{512kG5Mix}_{\text{BP}}$ had a 0.5\% mean absolute WER reduction compared to B2$^\text{Bytes}$, a byte baseline with a similar amount of decoder parameters. This again reveals the importance of the flexibility provided by the UML.

\section{Conclusions}
\label{sec:conclusions}
In this work, we develop the method of UML for multilingual ASR, which reuses each output node for a different subword unit for each language, making the output layer monolingual and language universal. A simple tokeniser-level change addresses many challenges in building massive-scale multilingual ASR such as the significant increase in output layer size when integrating more languages with one system.  Production scale experiments were conducted on 11 languages that justified the feasibility of building UML-based multilingual ASR systems with high accuracy. The potential of reducing the multilingual output size was also explored. When using a 4K-dim UML, the same averaged WER can be achieved compared to the baseline with a conventional 8K-dim WPM output layer. This reduces about 40\% of parameters in the decoder and considerably improves RNN-T training and test speed. Further reductions in size can be achieved by mixing WPMs with bytes using UML.

\vfill\pagebreak


\bibliographystyle{IEEEbib}
\bibliography{refs}

\end{document}